# Newton saw the truth — on the nature of fluid flow and viscous interaction

Jian He [a], Jin Wang [a], Qiaocong Kong [a], Penglong Zhao [a], Xiaoshu Cai [b], Xiaohang Zhang [c], Wennan Zou [a]✉.

[a] Institute of Fluid Mechanics/Institute for Advanced Study/College of Advanced Manufacturing, Nanchang University, Nanchang 330031, China; [b] School of Energy and Power Engineering, University of Shanghai for Science and Technology, Shanghai 200093, China; [c] School of Science, Nanchang Institute of technology, Nanchang, 330099, China.

*Wennan Zou.

**Email:** zouwn@ncu.edu.cn

## Abstract 1

Newton's thinking on the flow of viscous fluid comprise Book 2 of the Principia [1], which was extensively revised in the second edition, reflects his continued attention to the topic in his later years. His hypothesis that "the resistance arising from the want of lubricity in the parts of a fluid, is, other things being equal, proportional to the velocity with which the parts of the fluid are separated from each other" forms the conceptual foundation of viscous interaction. Since Stokes [2], the Newton's viscosity law is recast as the linear relation between the non-isotropic parts of viscous stress and the strain-rate, and so results in the Navier-Stokes (N-S) equations for the real viscous fluid [3]. However, the proposition of viscous force coming from stain-rate has never been experimentally demonstrated in a curved flow. Here we adopt the mechanism of slip instead of deformation to characterize fluid flow, and consequently formulate viscous friction in fluid. Following this assumption, we derive the velocity profile of the laminar Taylor-Couette (T-C) flow between two concentric cylinders in rotation, which confirm Newton's proposition in his ideal experiment, and especially design an experiment for the cylinders rotating at the same constant angular speed [4] to distinguish the new model from that underlying the N-S equations. We solemnly proclaim the experimental data do support the new model, which make clear the Stokes' viscosity stemming from deformation rate is not physically real. We believe the slip viscosity combined with new idea of continuum model will open a new routine to turbulence.

**Abstract 2**

A distinctive unambiguous simple experiment is designed to prove or disprove the viscosity model underlying the classical theory of fluid flow — the Navier-Stokes (N-S) equations. Though Newton once constructed the viscous friction coming from layering fluid, the mechanism finally proposed by Stokes that the viscous force is determined by strain-rate became the mainstream and so resulted in the N-S equations. Here we use modern mathematical tools to express the slip viscosity model in line with Newton's original idea, and deduce the laminar analytical solution of Taylor-Couette (T-C) flow between two concentric cylinders. When the outer cylinder is infinite, this solution reproduces the result of the ideal experiment proposed by Newton in the *Principle*, and shows a significant difference with the solution of the N-S equations when the diameter of the outer cylinder is relatively large. For this reason, we design a T-C flow experiment with inner and outer cylinders rotating with the same angular velocity, and the accurate measured data actually refute the strain-rate viscosity model. The slip viscosity model and the consequent new flow theory will greatly promote our conventional cognition on the evolution of continuous media, and provide a profound understanding of this ever open and closely connected world, which inevitably leads to a new vision in turbulence research.

**Main Text**

**1. Introduction**

Fluid flow is one of the basic processes in macroscopic physical phenomena. Viscosity is the main property of fluid in flow to measure the ability of fluid to resist flow. Viscosity is also essential for flow evolving into complex vortices and turbulence. Euler [5] introduced that "a perfect fluid can sustain no shearing force", but in fact, almost all fluid has viscosity. When a body moves through stationary ambient fluid or a flowing fluid bypasses a body, there must exist viscous friction applied on the body. Through the careful observation of fluid flow, Newton [1] proposed that the shear force between the fluid layers relatively slipping over each other is proportional to the speed difference, that is,

$$\tau = \mu \frac{\delta U}{\delta h}, \tag{1}$$

where $\mu$ is the dynamic viscosity of the fluid. Stokes [2] made an analogy with the deformation of elastic solids to understand the flow of viscous fluid, and treated the viscous friction as the stress responding to strain rate. Thus, the linear constitutive relation between the viscous stress and the strain rate finally results in the well-known N-S equations [6,7]. According to the deformation model, the viscous force formula expressed by strain rate in plate Couette flow (Fig. 1 bottom) is

$$\boldsymbol{\sigma} = \mu \frac{U}{h}(\mathbf{e}_1 da_2 + \mathbf{e}_2 da_1); \tag{2}$$

that means, besides the ordinary viscous friction (1), there is a shear force on the area element $da_1$ perpendicular to the slip surface $da_2$, however this part cannot be directly measured or verified by experiments, so the deformation understanding by regarding the change of fluid flow as a pure strain plus a rotation (Fig. 1 bottom) just comes from people's subjective judgement on viscous interaction.

Microscopically, Kreuzer [8] pointed out that viscosity is a contact interaction based on molecular diffusion and/or adhesion, viscous force is a kind of internal friction in the flowing fluid, and the flow can be understood as a slip process (Fig. 1 top). Zou [9,10] proposed that "the ideal viscous fluid cannot bear any shear force without slipping", such that the interior of fluid in motion is full of some complete or incomplete layered structures. When the fluid layers with different velocities slide across each other, the viscous force in the fluid keeps on the surface of slip. Therefore, the additional shear component in (2) is nonexistent in the slip model of viscous flow, but its effect doesn't appear until the flow becomes curved. In this paper, the simplest curved flow, namely the Taylor-Couette (T-C) laminar flow between concentric cylinders [4,11,12] will be investigated, where the streamlines are all circular.

## 2. Theoretical solutions based on deformation and slip models

Consider the flow of fluid between two concentric cylinders which are upright standing and infinitely long, with radius $R_1$ and $R_2$, and rotating around their axis with angular speed $\Omega_1$ and $\Omega_2$, respectively. The fluid is assumed to be incompressible, its flow is laminar. So, under the cylindrical coordinate system ($r$, $\theta$, $z$) with the axis of cylinders as $z$-axis, we get the flow fields with properties

$$v_r = v_z = 0, \tag{3}$$

$$v_\theta = v_\theta(r), p = p(r, z), \tag{4}$$

and satisfying the boundary conditions

$$v_\theta = \Omega_1 R_1, r = R_1; v_\theta = \Omega_2 R_2, r = R_2. \tag{5}$$

The analytic solution from the N-S equations [12] can be solved to be

$$V_{\text{N-S}} = v_\theta = \frac{\Omega_2 R_2^2 - \Omega_1 R_1^2}{R_2^2 - R_1^2} r + \frac{\Omega_2 - \Omega_1}{R_2^{-2} - R_1^{-2}} r^{-1}. \tag{6}$$

When the two cylinders rotate at the same constant angular speed $\Omega_1 = \Omega_2 = \Omega$, the velocity (6) becomes

$$V_{\text{N-S}} = \Omega r, \tag{7}$$

that means the fluid between the cylinders admit a rigid-body rotation. Is it true? Has it been confirmed or falsified by experiments. Let's face it.

Before doing so, we present a different theoretical analysis when the flow is regarded as *slip* between fluid layers with molecular scale. Looking back on history, Newton's elaboration on the mechanism of fluid viscous friction is definitely based on the conjecture that stratified fluids slip over each other, where the interface is specially emphasized not to be plane. Newton even constructed an ideal experiment, that is, the steady flow caused by an infinitely long cylinder constantly rotating around its axis in infinite fluid, and concluded that the fluid will rotate around the cylinder at a constant speed. But the solution (the Principia Book 2 Section 9 Corollary 1) [1] is inconsistent with $\Omega_1 R_1^2 r^{-1}$ coming from the N-S equations. Smith [13] argued that "Newton mistakenly balanced the forces on the inside and outside of each thin fluid shell surrounding the body, not the torques". That is to say, the deformation model underlying the N-S equations doesn't support the Newton's attention.

The slip structure is a kind of tension state of fluid layering formed in response to non-uniformity of flow [10]. The physical effect of slip structure, coupling with flow and vortex evolution, dwells in the difference and nonintegrability. We call the difference of slip structure the swirl field, which is usually independent of but coupled with the velocity field [9,10]. In the case of steady laminar flows, the slip structure of fluid is completely determined by the streamline. Introduce the Cartesian coordinate system with basis $(\mathbf{e}_1, \mathbf{e}_2, \mathbf{e}_3)$, the velocity can be expressed with its direction angles $(\varphi, \phi)$ as

$$\boldsymbol{v} = v_i \mathbf{e}_i = V\mathbf{n}(\varphi, \phi) = V(\mathbf{e}_1 \cos\varphi \sin\phi + \mathbf{e}_2 \sin\varphi \sin\phi + \mathbf{e}_3 \cos\phi), \tag{7}$$

where $V$ is the speed. Then, the swirl field, as an axial-vector valued 1-form, can be characterized by

$$A_k^i \mathbf{e}_i dx_k \equiv -\mathbf{n} \times d\mathbf{n} = -\varepsilon_{ilm} \mathbf{e}_i n_l dn_m. \tag{8}$$

Any alternative definition of swirl field will change the evaluation of streamwise vortex, but doesn't affect its coupling with the velocity. Thus, the Newton's thinking on fluid with linear constitutive relation can be indicated by

$$\sigma_{ji} = \mu D_j v_i \equiv \mu(\partial_j v_i + \varepsilon_{ikl} A_j^k v_l) = \mu(\partial_j V) n_i, \tag{9}$$

where $\mu$ is the dynamic viscosity of the fluid, and the covariant derivative $D$ is carried out through the *swirl field* as the connection. Further, the resultant of viscous friction can be calculated by

$$D_j \sigma_{ji} \equiv \partial_j \sigma_{ji} + \varepsilon_{ikl} A_j^k \sigma_{jl} = \mu(\nabla^2 V) n_i, \tag{10}$$

and finally, we reach the controlling equations of steady laminar flows as

$$\rho v_j \partial_j v_i = -\partial_i p + \mu(\nabla^2 V) n_i, \tag{11}$$

besides the continuity equation $\partial_k v_k = 0$. Introducing the vorticity $\Omega^i = \varepsilon_{ijk} \partial_j v_k$ and the total pressure $P = \frac{p}{\rho} + \frac{1}{2}V^2$, (11) can be rewritten as

$$\partial_i P = -\varepsilon_{ijk} \Omega^j v_k + \nu(\nabla^2 V) n_i, \tag{12}$$

which means, under the slip model, the gradient of total pressure can be decomposed into two orthogonal parts: the streamwise part to balance the viscous force and the part perpendicular to the stream to balance the centrifugal force of the bending flow. Based on the above equations of the slip model, Zhu [14] used the UDF technique of software FLUENT to realize the numerical simulation of laminar flow, while Xu [15] carried out numerical analyses and comparison between the classical model and the slip model, including the T-C flow between concentric cylinders with ends.

Now back to the laminar T-C flow with $\mathbf{n} = \mathbf{e}_\theta$, it is easy to derive that $A_k^i \mathbf{e}_i dx_k = -\frac{\mathbf{e}_z}{r} d\theta$, and the controlling equation of the velocity becomes

$$\nabla^2 v_\theta = \frac{1}{r}\frac{\partial}{\partial r}\left(r\frac{\partial v_\theta}{\partial r}\right) = 0. \tag{13}$$

Using the boundary conditions (5), we get the analytic solution

$$V_{\text{new}} = v_\theta = \Omega_1 R_1 + (\Omega_2 R_2 - \Omega_1 R_1)\frac{\ln r - \ln R_1}{\ln R_2 - \ln R_1}. \tag{14}$$

It degenerates into Newton's solution by setting $R_2 \to \infty$ and $\Omega_2 R_2 = \Omega_1 R_1$. Since the water particles are statistically in circular motion, we introduce the orbital angular speed by $\omega = V(r)/r$. Then when $\Omega_1 = \Omega_2 = \Omega$, we have for the new model

$$\frac{\omega}{\Omega} = \frac{V_{\text{new}}}{V_{\text{N-S}}} = \frac{R_1 + (R_2 - R_1)\frac{\ln r - \ln R_1}{\ln R_2 - \ln R_1}}{r} = \frac{\ln(1+\eta) + \eta\ln(1+\delta\eta)}{(1+\delta\eta)\ln(1+\eta)}, \tag{14}$$

where the dimensionless cylinder spacing (DCS) $\eta \equiv (R_2 - R_1)/R_1$, and the relative radial distance (RRD) $\delta \equiv (r - R_1)/(R_2 - R_1)$, as shown in Fig. 2. It is easy to prove that the maximal value of $\omega$ is a monotone increasing function of $\eta$. For example, when $\eta = 0.3$, the position with largest difference is at $\delta = 0.4455$, while the maximum ratio between the slip model and the classical model is 1.0086; but when $\eta = 0.8$, the position with largest difference is at $\delta = 0.3797$, while the maximum ratio between the slip model and the classical model is $1.0439$. In order to distinguish the deformation model and the slip model, we draw the velocity profile with the DCS $\eta = 0.3, 0.6$ and $0.8$ in Fig. 2, where the abscissa $\delta \in [0,1]$, and the ordinate is scaled by the N-S solution $V_{N-S} = \Omega r$, say $V/V_{N-S}$. It is obvious from Fig. 2 that the largest difference happens near the center of radial position, and as the DCS increases the relative difference of velocity becomes larger, and in the meantime the position with the largest difference gradually approaches the inner cylinder.

The difference near one percent or more is unlikely to be an experimental error, which pushes us about ten years ago to carry out an experimental study with accurate measurement on the velocity profile for the laminar T-C flow, which is now available as shown in Fig. 3. The measured velocity profile or orbital rotation map (Fig. 4) of water flow confined between two concentric cylinders rotating as a rigid-body definitely support our prediction (14) according to the new model.

## 3. Conclusion Discussion

The experimental results reported here falsify the viscous friction model underlying the N-S equations, and will greatly change our understanding of complex flows, including turbulence. We may conclude through the experiment that: (1) the N-S equations cannot correctly predict the velocity profile of curved flows, let alone the detail of turbulent flows; (2) the deviation of the prediction by the N-S equations from the real results increases with the shear thickness scaled by the mean radius of curvature of the streamlines; (3) Newton saw the truth, but most people misused his proposition, since the slipping fluid layers described in the *Principia* is obviously not limited to be of plane; (4) on the contrary, the slip model, as a phenomenological model, accurately predicts the experimental results of the laminar T-C flow. Inspired by the slip viscosity model, which conforms to Newton's original idea, but needs to be expressed with the language of modern differential geometry, namely using an additional swirl field as the spatial connection of velocity field, we make an important statement that *the nature of flow is the slip of fluid layered at molecular scale, instead of the deformation induced by the inhomogeneous displacement of continuously aggregative fluid particles*, which implies that the viscous force is determined by the slip strength between fluid layers, instead of the strain-rate, too. This novel understanding can be related to the microscopic mechanism of viscosity of fluids, whatever gas or liquid.

**Acknowledgments:** The authors thank Prof. Chao Sun of Tsinghua University for helpful discussions, and thank to all the people who have provided help and encouragement in the process of this research. W.N.Z. acknowledges the support of the Department of Engineering Mechanics, Nanchang University, especially the support of Prof. Chun Zhang, the dean of the department.
**Data availability:** The data that support the findings of this study are available from the corresponding authors on reasonable request.
**Code availability:** The current reconstruction codes used in this study are available from the corresponding authors on reasonable request.

**Figures and Tables**

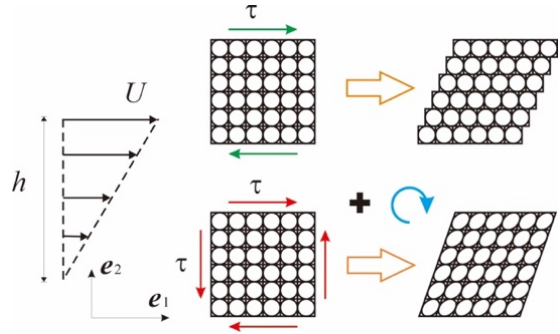

**Figure 1.** Slip (top) and deformation (bottom) understandings of the Couette flow between two parallel plates.

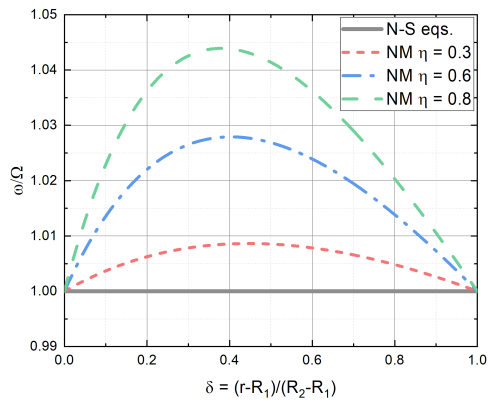

**Figure 2.** The orbital angular speed $\omega$ of new model ('NM' in the figure) scaled by the N-S solution with different relative cylinder spacing (RCS) $\eta \equiv (R_2 - R_1)/R_1$.

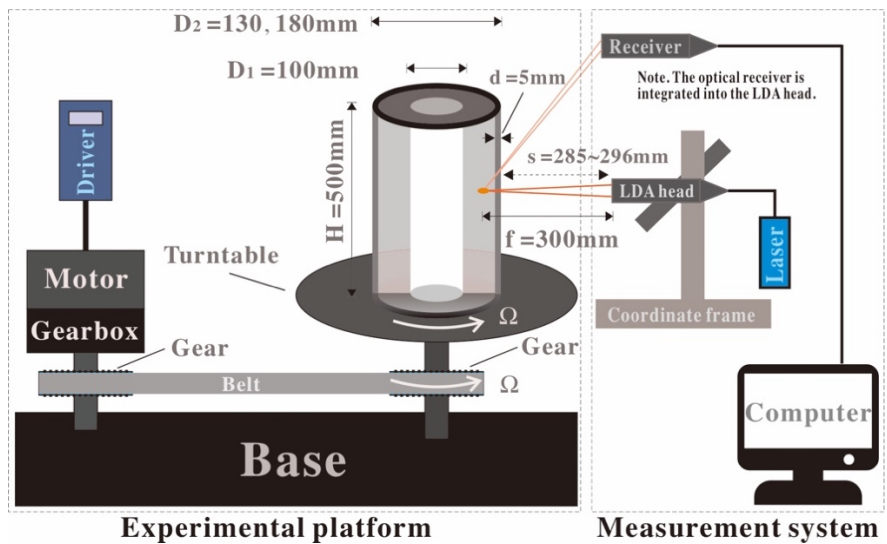

**Experimental platform**          **Measurement system**

**Figure 3.** Schematic diagram of experimental platform and measurement system.

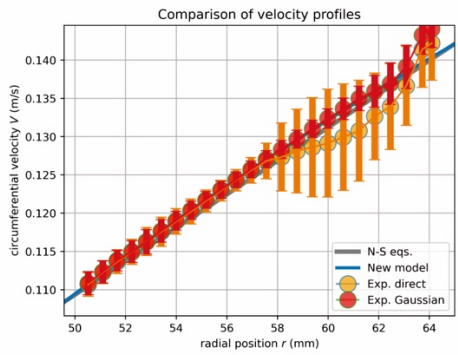

(*a*) velocity profile for $D_2 = 65$mm

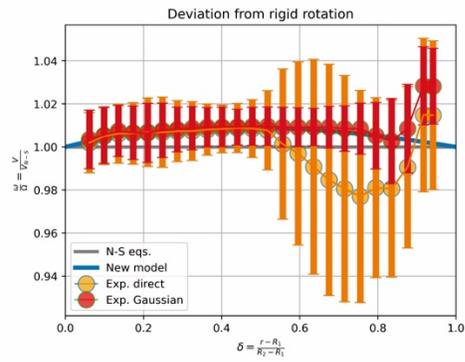

(*b*) scaled orbital angular speed for $D_2 = 65$mm

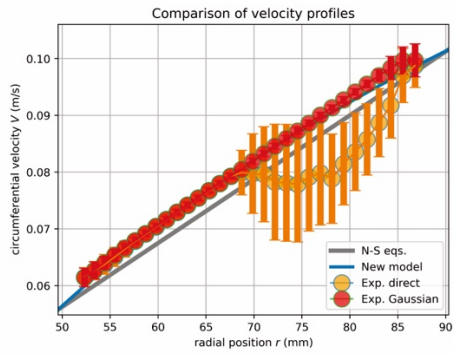

(*c*) velocity profile for $D_2 = 90$mm

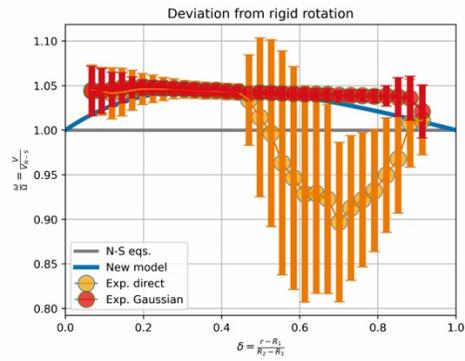

(*d*) scaled orbital angular speed for $D_2 = 90$mm

**Figure 4.** Experimental results show that the N-S equations cannot predict the velocity in the laminar T-C flow: (*a*)(*c*) velocity profile and (*b*)(*d*) scaled rotation speed for experimental parameters $R_2 = 65$mm, $\Omega = 20.88$rpm and $R_2 = 90$mm, $\Omega = 10.74$rpm, respectively.

## Materials and Methods

### 1. Experiment setup

As shown in Fig. 3, the concentric cylinder frame used for T-C flow experimental platform for rigid body rotation was built by Wang Jin [16] under the guidance of Professor Cai Xiaoshu of University of Shanghai for Science and Technology in 2019, while the LDA measurement system is financially supported by Institute of Fluid Mechanics, Nanchang University. He Jian improved the coordinate frame under the guidance of Zhao Penglong in 2022, and improved the precise control ability of the laser head to emit laser under the supervision of Zou Wennan in 2023. The parameters of LDA measurement system are listed in Table 1.

In our experiment, the machining accuracy of the mechanical part is extremely high. For example, the levelness of the base is within 1 wire (100 wires =1mm), the co-axiality of the rotating shaft part connecting the rotary table is within 3 wire (0.03mm), all places in direct contact with water are chrome plated for rust prevention, and the hole matching error is within 0.1mm.

### Flow environment

The concentric cylinders used in the experiment are made of plexiglass (acrylic PMMA), with high mechanical strength and light transmittance up to 96%. The inner cylinder diameter is $D_1 = 100$mm, the outer cylinder diameter could be $D_2 = 130$mm or 180mm, the wall thickness of the outer cylinder is $d = 5$mm, while the cylinder height is $H = 500$mm. The water between cylinders moves with the rigid rotation of turntable, which is driven by a belt driving system with toothed pulleys, installed on a very stable base. This belt system does not change the rotation speed from the active end. The diameter and height of the cylinders are claimed to have errors of ±0.2mm when leaving the factory. After cutting, grinding and polishing, the circular runout of the solid inner cylinder is promised to be within 3 wires (0.03mm), while all height errors are within 0.1mm, but it is quite difficult to further process the hollow outer cylinder.

Unfortunately, we found significant deviations from the above expected values after carefully geometrical measurement of cylinders, as shown in Table 2. It is obvious that the mean values are not the preset ones and the ranges exceed the promised errors, respectively. And it is learned that the precision machining of acrylic glass cylinders is extremely difficult. The non-ideality in manufacturing has led to distortion of the measurement signals, forcing us to develop new analytical tools for speed data screening. Particularly, (1) the radius of the inner cylinder after secondary processing has a smaller range, less than 0.1mm, but its average also differs significantly from the predetermined value 5mm; (2) the ranges of thickness of outer cylinders are about 0.6mm while the means of thickness deviate from the expected values; (3) the exterior diameters of outer cylinders have ranges larger than 0.8mm; (4) the cylindricities are also not ideal.

### Power and control

The active end of the belt driving system is first connected with a planetary gearbox (Type: PLF120-10-S2) as retarder in the gear ratio 10:1 and next driven by an AC servo motor (Model: 130TS-M06025, with power 1.5kW, torque 6N·m and rotated speed 2500rpm), which is controlled by an AC servo driver (Model: TD-AB SF980, used for 1500W servo motor at AC 3PH 220V).

### End effect

It is impossible to carry out a T-C flow experiment with infinite long cylinders, as reported by the theoretical model. In practice, we have to build up an experiment platform with finite length, but need to choose the length of cylinders to guarantee the velocity profile of some cross sections not affected by the ends. Under the viscosity model of the N-S equations, the fluid moves as the rigid rotation of the ends, so completely independent of the length of the cylinders. When analyzed according to the slip viscosity model, numerical calculation shows the scale of effect range by the end is about the same as the cylinder spacing [17]. Thus, we adopt $H = 500\text{mm}$ which is enough for all involved outer cylinders to confirm that the velocity profile of most cross sections near the middle are almost the same as the theoretical solution without ends.

**Measurement system**

In this experiment, the Laser Doppler Anemometer/Velocimeter (LDA/LDV) of Dantec Company is used to measure the velocity profile of the T-C flow between two concentric cylinders, as shown in Fig. 3. LDA is a high-precision, single point, non-contact technology for measuring fluid velocity, which requires adding tracer particles to the tested flow fluid. The tracer particles used in this experiment are hollow glass beads with a density of $1.03\text{g/cm}^3$, a refractive index of $1.5$ and an average nominal diameter of $10\mu\text{m}$, which can offer good scattering efficiency and a sufficiently small velocity lag. The specific parameters of the LDA system used in this experiment are shown in Table 2.

Good mixing of tracer particles in water flows is required. Experience has shown that any powders introduced into a liquid flow as seeding should first be well mixed in a smaller volume of liquid and then poured in. This can greatly prevent agglomeration or immediate sedimentation or adhesion to surfaces. According to Durst [18], for the ellipsoidal detection volume with three radii $a_d = 0.04873\text{mm}$, $b_d = 0.04848\text{mm}$ and $c_d = 2fa_d/h = 0.4873\text{mm}$, by assuming the mean number of particles simultaneously in the volume to be $0.1$, the maximal particle concentration is $n_p < \frac{0.1}{V_d} = 20.74\text{mm}^{-3}$, which means particles are separated on average by a distance of $364\mu\text{m}$. One particle with $10\mu\text{m}$ diameter having mass of $5.393 \times 10^{-7}\text{mg}$, this leads to at most $11.19\text{mg}$ particles in a liter of water. On the other hand, Albrecht *et al* [19] pointed out that the minimal particle concentration depends on the integral time scale $T_u < 10\text{ms}$, the mean velocity $\bar{u} < \Omega R_2 = 0.1886\text{ms}^{-1}(\Omega = 20\text{rpm})$ across the interference plane $A_d = \pi b_d c_d = 0.0742\text{mm}^2$ in the measurement volume, and so has a value $n_p > \frac{0.5}{T_u\bar{u}A_d} > 3.573\text{mm}^{-3}$, or $1.927\text{mg}$ particles in a liter of water. In practice, the seeding amount we use is close to the lower particle concentration.

According to the analysis of the Doppler effect in the light propagation, two light waves intersect at an angle $\theta$ (MM Fig. 1), and form a cross area called the measurement volume. A tracer particle suspended in the flow scatters the lights of two laser beams simultaneously when it passes through the measurement volume. Because of the different spatial layout of two laser beams, the moving particle perceives the different light frequencies resulted from different Doppler effects. While being received by the photodetector, the frequency difference of two superimposed light waves, which is called the Doppler frequency, can be related to the component $v_\perp$ of the particle velocity perpendicular to the bisector of the two laser beams as

$$v_D = \frac{2v_\perp}{\lambda} \sin\frac{\theta}{2},\tag{S1}$$

where $\lambda$ is the wavelength. From the fringe model on the light interference, the light intensity alternates with a distance equal to

$$\Delta x = \lambda / \sin\frac{\theta}{2},\tag{S2}$$

which is known as the fringe spacing in the measurement volume, and is easy to be measured. In our experiment, the first-dimension laser along the radial direction is used to measure the circumferential velocity of the T-C flow, and so $\lambda =$ 660 nm and $\tan\frac{\theta}{2} = \frac{h}{f} =$ 0.05 $\Longrightarrow \frac{\theta}{2} =$ 0.04996.

**Pressure leakage, and bubble control**

After pouring a small portion of water sample dissolved with tracer particles into a large water body, stirring is required to make the mixture uniform. It can be considered that supersaturated air has also been dissolved in the whole water body. When the water is poured into the cylinder cavity, if the upper end surface is sealed for testing only after the separated bubbles are only discharged during standing still for hours, the bubbles in the water will continue to precipitate and converge to the upper end surface due to the temperature rise and the pressure change inside the fluid (pressure reduction in some water areas). This disturbs the boundary constraints of the flow, resulting in pressure leakage and failure to achieve the desired flow. In order to control the generation of bubbles during the flow experiment, we have adopted two measures: (1) removing bubbles and sealing the ends when the water body is at a higher temperature, and conducting the experiment when the water body is at a lower temperature; (2) because the pressure change caused by the flow is proportional to the square of the rotational speed, using a smaller rotational speed can significantly reduce the range of pressure changes in the water within the cylinders.

## 2. Correction, measurement and preliminary treatment of data

**Optical analysis and correction**

In our experiment, as shown in MM Fig. 1, the intersection of two laser beams becomes the point B instead of the point A, and the angle $\theta_\mathrm{w}$ of intersection is not equal to $\theta_\mathrm{a} \approx$ 5.711°, but the reported velocity by the measurement system is calculated according to the measured $v_D$ and the parameters $\theta_\mathrm{a}$ and $\lambda_\mathrm{a}$ of the air, that means we need to revise the velocity $v'_\perp = v_D \Delta x_\mathrm{w} = k v_D \Delta x_\mathrm{a} = k v_\perp$ and the position of the measurement volume changes from $r_A = R_2 + d + s - f$ to $r_B = R_2 - \left[h - \frac{\theta_\mathrm{a}}{2}s - d(\alpha_{\mathrm{g}1} + \beta_1)\right]\mathrm{ctan}\frac{\theta_\mathrm{w}}{2}$.

In the following, the refractive indices of air, water and glass are taken to be

$$n_\mathrm{a} = 1, n_\mathrm{w} = 1.333, n_\mathrm{g} = 1.491, \tag{S3}$$

respectively. When all incident and refraction angles are assumed to be small so that for any involved angle $\alpha$ we have the following approximate treatment

$$\sin\alpha \approx \tan\alpha \approx \alpha. \tag{S4}$$

From the Snell law of refraction,

$$n_\mathrm{a}\sin\alpha_\mathrm{a} = n_\mathrm{g}\sin\alpha_{\mathrm{g}1}; \; n_\mathrm{g}\sin\alpha_{\mathrm{g}2} = n_\mathrm{w}\sin\alpha_\mathrm{w}, \tag{S5.1}$$

and the geometrical relations

$$\alpha_\mathrm{a} = \frac{\theta_\mathrm{a}}{2} - \beta_1, \qquad \alpha_{\mathrm{g}2} = \alpha_{\mathrm{g}1} + \beta_1 - \beta_2, \qquad \frac{\theta_\mathrm{w}}{2} = \alpha_\mathrm{w} + \beta_2, \tag{S5.2}$$

one can derive

$$\beta_1 = \frac{h - \frac{\theta_\mathrm{a}}{2}s}{R_2 + d}, \beta_2 = \left(1 + \frac{n_\mathrm{a}}{n_\mathrm{g}}\frac{d}{R_2}\right)\beta_1 - \frac{n_\mathrm{a}d}{n_\mathrm{g}R_2}\frac{\theta_\mathrm{a}}{2}, \tag{S5.3}$$

and the velocity modification coefficients $k$ and the real position $r_B$ of measuring point have the expressions

$$k = \left[1 + \frac{n_\mathrm{g} - n_\mathrm{w}}{n_\mathrm{g}}\frac{d}{R_2} + \frac{f - s}{R_2 + d}\left(\frac{n_\mathrm{w} - n_\mathrm{a}}{n_\mathrm{a}} - \frac{n_\mathrm{g} - n_\mathrm{w}}{n_\mathrm{g}}\frac{d}{R_2}\right)\right]^{-1}, \tag{S6}$$

$$r_B = R_2 - \frac{\left(R_2 + \frac{n_a}{n_g}d\right)\frac{f-s}{R_2+d} - \frac{n_a}{n_g}d}{\frac{R_2+d}{R_2}\frac{n_a}{n_w} - \frac{n_a}{n_g}\frac{d}{R_2} + \left(1 + \frac{n_a}{n_g}\frac{d}{R_2} - \frac{R_2+d}{R_2}\frac{n_a}{n_w}\right)\frac{f-s}{R_2+d}}. \tag{S7}$$

As comparison, the position without modification is $r_A$. In practice, 24 points controlled by the parameter $s$ are chosen as the measurement points (MPs), where the real radial coordinates and the modifying coefficients of circumferential velocity for them are shown in MM Fig. 2, calculated from (S6) and (S7).

Without the assumption of small angles, the geometric relation

$$[f - s - (R_2 + d)(1 - \cos\beta_1)]\tan\frac{\theta_a}{2} = (R_2 + d)\tan\beta_1, \tag{S8}$$

Can be used to derive a recursive process $\tan\beta_1 = \left(\frac{f-s}{R_2+d} - 1 + \cos\beta_1\right)\tan\frac{\theta_a}{2}$ starting from $\beta_1 = \frac{f-s}{R_2+d}\frac{\theta_a}{2}$ to get $\beta_1$. Then following the explicit relations

$$\alpha_a = \frac{\theta_a}{2} - \beta_1, \alpha_{g1} = \sin^{-1}\left(\frac{n_a}{n_g}\sin\alpha_a\right) \tag{S9}$$

we can use the geometric relation

$$\frac{(R_2+d)\sin\beta_1 - R_2\sin\beta_2}{(R_2+d)\cos\beta_1 - R_2\cos\beta_2} = \tan(\beta_1 + \alpha_{g1}), \tag{S10}$$

to obtain the second recursive formula $\sin\beta_2 = \left(1 + \frac{d}{R_2}\right)\left[\sin\beta_1 - \cos\beta_1\tan(\beta_1 + \alpha_{g1})\right] + \cos\beta_2\tan(\beta_1 + \alpha_{g1})$ starting from $\beta_2 = \beta_1 - \frac{d}{R_2}\alpha_{g1}$ to get $\beta_2$. More subsequent parameters are calculated from

$$\alpha_{g2} = \alpha_{g1} + \beta_1 - \beta_2, \alpha_w = \sin^{-1}\left(\frac{n_g}{n_w}\sin\alpha_{g2}\right), \frac{\theta_w}{2} = \alpha_w + \beta_2. \tag{S11}$$

Finally, we calculate the modification coefficient $k$ of velocity and the real position $r_B$ of measuring point by

$$k = \frac{n_a\sin\frac{\theta_a}{2}}{n_w\sin\frac{\theta_w}{2}}, r_B = R_2\cos\beta_2 - R_2\sin\beta_2\text{ctan}\frac{\theta_w}{2}. \tag{S12}$$

The numerical practice shows that all relative errors of $k$ and $r_B$ are below 0.06% for both $R_2 = 65\text{mm}$ and 90mm, as Zhang[18] pointed out. The correction parameters $k$ and $\Delta r = r_A - r_B$ are shown in MM Fig. 2.

**Gaussian distribution of velocity samples**

In our LDA measurement, the waist diameter of two pairs of laser beams before passing through the front lens is $d_l = 2.6\text{mm}$, according to the calculation formula, the waist diameter of two laser beams intersecting to form a measurement volume is

$$d_f = \frac{4f\lambda}{\pi E d_l}, \tag{S13}$$

equal to 0.09696mm when $f = 300\text{mm}, \lambda = 660\text{nm}$ are used for measuring the circumferential velocity component and the beam expansion factor $E = 1$. Considering the half included-angle of the laser beams $\frac{\theta_a}{2} = 5.711°$ or $\tan\frac{\theta_a}{2} = 0.1$, we can get the radial scale of the measurement volume to be $2a = d_f/\sin\frac{\theta_a}{2} = 0.974457\text{mm}$. Since the laser intensity has a Gaussian distribution (cf. Zhang[18]) along the transverse direction, the tracer particles appearing in the center of the measurement volume will cause the largest number of signals to trigger, so the different flow velocities of the fluid within the range of the radial spread $2a$ of the measurement volume will also show a Gaussian distribution characteristics, with a relative velocity difference range about $\frac{\Delta V}{V} = \frac{a}{V}\frac{dV}{dr} \approx \frac{a}{r} \in [0.0075, 0.0097]$ for $D_2 = 130\text{mm}$ or $[0.0054, 0.0097]$ for $D_2 = 180\text{mm}$, which may need a small correction because the two lasers actually intersect in water, and the intersection angle is smaller than that analyzed in air. As Zhang (2010) [18] reported, the flow fluctuations measured by LDA, even in turbulent flows, should approximately fulfill the Gaussian probability distribution.

**Measurement process and preliminary treatment**

Before the test, we mixed the $10\mu\text{m}$ tracer particles evenly by stirring, which can be seen from the signal received in less than two milliseconds at some measurement points. Due to the finiteness of the laser beams and the measurement volume, the limit range of control position is $s \in [284.5\text{mm}, 294.8\text{mm}]$, or equivalently $r_B \in [50.5\text{mm}, 64.0\text{mm}]^{\clubsuit}$, smaller than the theoretical range $r \in [50\text{mm}, 65\text{mm}]$. When the electronic gear ratio of the AC servo driver is adjusted to $3:1$, a $50\text{ kHz}$ controller can be used to achieve $0$-$90$ rpm cylinder rotation, and the stability deviation is less than $0.2\%$. We conduct alignment inspection before startup, and start to collect $2 \times 10^4$ consecutive samples from each measurement point after the cylinders rotate for one hour or later, depending on that the flow in the central section is stable. We control the signal quality by using LDA monitoring parameters "Sensitive" and "Gain" (see the supplementary material "Test Log") to measure the circumferential velocity component of $24$ points along the radial direction accordingly; then, continuously measure the rotating angular speed for $100$ times as the working condition parameter of this measurement.

For the obtained raw data, we first do the following preliminary treatment:

(1) Remove the sample whose value is near zero (it is considered as the sampling triggered by the scattering light of the cylinder wall);

(2) Then, correct the speed of each measurement point with the velocity correction coefficient given by the formula (S6) as shown in MM Fig. 2;

(3) Using $v_{\min} = \Omega R_1$, $v_{\max} = \Omega R_2$ for reference, eliminate the abnormal samples with values less than $v_{\min}/1.2$ or larger than $1.2v_{\max}$, which are believed to be caused by molecular Brownian motion and other factors.

(4) For every measurement point, calculate the average value $v_{\text{m}}$ and standard deviation $\Delta_v$ from the sieved samples, and use the values $v_{\text{m}} \pm 5\Delta_v$ to screen the data again.

The remaining signals, say for example the signal segments with $D_2 = 130\text{mm}, H = 0\text{mm}, \Omega = 20.88\text{ rpm}$ are shown in MM Fig. 3.1-4 while the average velocity profile and the orbital angular speed scaled by $\Omega$ are illustrated in Fig. 4(*a*)(*b*) with orange dots (lines), will be analyzed carefully.

**3. Reflection, contamination and extraction of velocity profile**

**Magnitude spectrum and probability distribution of velocity samples**

Based on the $2 \times 10^4$ consecutive signals at every measurement point, besides extracting a signal segments, we consider the signals as a time sequence and calculate the spectrum of the magnitude scaled by the local average speed, and count out the probability distribution of velocity samples by regarding the signals as random samples, as shown in MM Fig. 3.1-4. In addition, among the probability distribution of velocity of each MP, the independent and narrower peak is fit by a Gaussian distribution, as shown in MM Fig. 3.1-4 and MM Fig. 7.1-6.

---

$^{\clubsuit}$ Since the optical correction depends on the geometrical parameters $\{R_1, R_2, d\}$, which are shown in Table 2 to be different to some extent at different cross-sections, we will proceed the data treatment in the following, as an example, using the expected parameters, namely $\{R_1 = 50\text{mm}, R_2 = 65\text{mm}, d = 5\text{mm}\}$. But in the result displays, such as Fig. 4, MM Figs. 4-6, the analyses will be carried out according to the actual measured geometric parameters of different cross-sections.

**Basic features of velocity data**

Three features of the measured data should be mentioned: (1) the characteristics of cycle period is obvious; (2) the velocity of MPs near the walls unreasonably becomes higher, and the mean square error (MSE) is larger than that of the expected Gaussian distribution; (3) from the middle of inner and outer cylinder walls to outer cylinder wall, agglomeration peak with low speed appears and becomes stronger, and gradually integrates into one with the Gaussian peak of the preset MP since it moves faster than the later. In addition, for the larger outer cylinder with $D_2 = 180\text{mm}$, the agglomeration peak with low speed could become the main one with higher weight, as shown in Figs. 7.5-6.

**Light reflection of moving surfaces and contamination due to fine structures**

It is difficult to provide a complete and thorough explanation for the above features. Two extra factors must be considered: the light reflection from the moving surfaces of cylinders, and the effects of fine structures on the surfaces that disrupt their roundness and cylindricity. The later include disturbing the flow and causing the dispersion of laser beams and the oscillation of laser focus. And further, the dispersion of laser beams will bring more complex reflection of light. Therefore, it is inevitable that the obtained signals are distorted by the fine structures and contaminated by the reflection of moving walls, especially when the MPs are close to the inner or outer cylinder walls.

**Story of signal changes**

First of all, the characteristics of cycle period can be explained by the irregularity of fine structures, but the features of velocity probability distribution changing with the radial position are not easy to understand. In order to tell a whole story, we need to make the following assumptions.

(1) The disturbance to the flow due to the fine structures is neglectable, especially for the MPs far from the wall surfaces.

(2) The radial oscillation of the laser focus may increase the standard deviation of velocity, but is generally negligible.

(3) The beam dispersion decreases the standard deviation of velocity from the main optical path and forms some secondary optical path: (i) for the MPs in the side near the inner cylinder wall, only the signals of the main optical path are sensible by the receiver, (ii) for the MPs in the side near the outer cylinder wall, a low-speed agglomeration of the Doppler signals of the secondary optical path is gradually detected by the receiver, moves faster than the MP towards the outer cylinder wall, and gradually increases its signal weight, and ultimately merges with the signal peak of the main optical path.

(4) The effect of light reflection from the moving wall surfaces becomes serious when the MP is close to the cylinder walls, complex and unstable when the MP is very close to the walls, resulting in a larger magnitude of velocity with a wider standard deviation, and more obvious cycle periodicity near the outer cylinder. Comparing the results of different outer cylinders, it is clear that the lifting of velocity magnitude due to the light reflection is greater for the larger radius $R_2$.

Based on the above analyses, the characteristics of signals of the stable laminar TC flow with inner and outer cylinders rotating at the same constant angular speed, as shown in Fig. 4, or detailed in MM Fig. 3 and MM Fig. 7, can be understood.

**Extraction of velocity signals**

Following the viewpoint that the real signals by the LDA signals should be of Gaussian distribution, and the normalized standard deviation should be less than 0.0097, we have developed the following data treatment method to recover the real velocity profile from the obtained data as much as possible. First, we find the independent and narrower peak of the velocity distribution obtained from all samples, and fit the peak with a single Gaussian distribution. Then, the mean of the distribution can be taken to be the velocity of preset point, as shown in MM Fig, 3 and MM Fig. 7 for examples; (2) the velocity values with standard deviations less than 0.0097 are considered to be reliable, say MPs 9-17 ($r = 55.76 - 59.99\text{mm}$) for $D_2 = 130\text{mm}$ and MPs 10-31 ($r = 57.61 - 81.79\text{mm}$) for $D_2 = 180\text{mm}$, which can be used to prove or disprove different viscosity models, and in Fig. 4 two typical results are illustrated.

**4. Original database**

As shown in Fig. 4, the fishbone feature of velocity profile or the arched feature of orbital rotation speed is universal for all test conditions.

For the radius $R_2 = 65\text{mm}$, after one-hour and three-hour rotating, we have measured the velocity profiles with rotation speeds $\Omega = 20\text{rpm}, 35\text{rpm}, 45\text{rpm}$ and $60\text{rpm}$, and/or at cross-sections $H = 0\text{mm}, 20\text{mm}$ and $50\text{mm}$ from the center. In MM Figs. 4-6, the Gaussian distribution weight, standard deviation of extracted velocity, and orbital angular speed for different parameters are compared. If no otherwise specified, the measurement begins at one hour after startup.

For the radius $R_2 = 90\text{mm}$, the pressure variance along the radial direction becomes higher quickly at the same rotation speed. This places higher demands on the closure of the upper end of the cylinders, and as a result, it is prone to flow instability. So, we have measured the velocity profiles after two-hour and four-hour rotating, with rotation speeds $\Omega = 10\text{rpm}$ and $13\text{rpm}$ and/or at cross-sections $H = 0\text{mm}$ and $50\text{mm}$ from the center. In MM Figs. 8-10, the Gaussian distribution weight, standard deviation of extracted velocity, and orbital angular speed for different parameters are compared. If no otherwise specified, the measurement begins at two hours after startup.

In

**Figures and Tables for Materials and Methods**

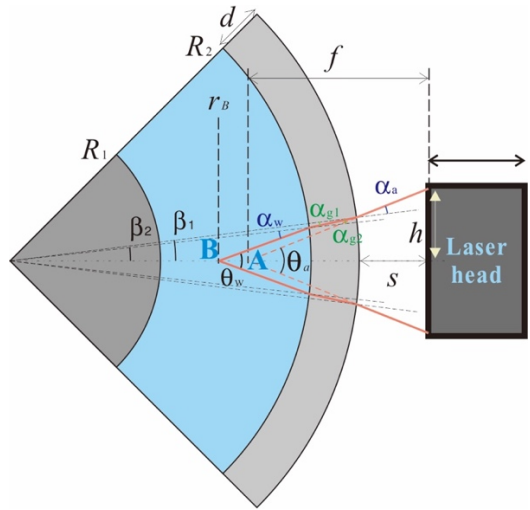

**MM Figure 1.** The optical paths for the measurement of the circumferential velocity in our experiment.

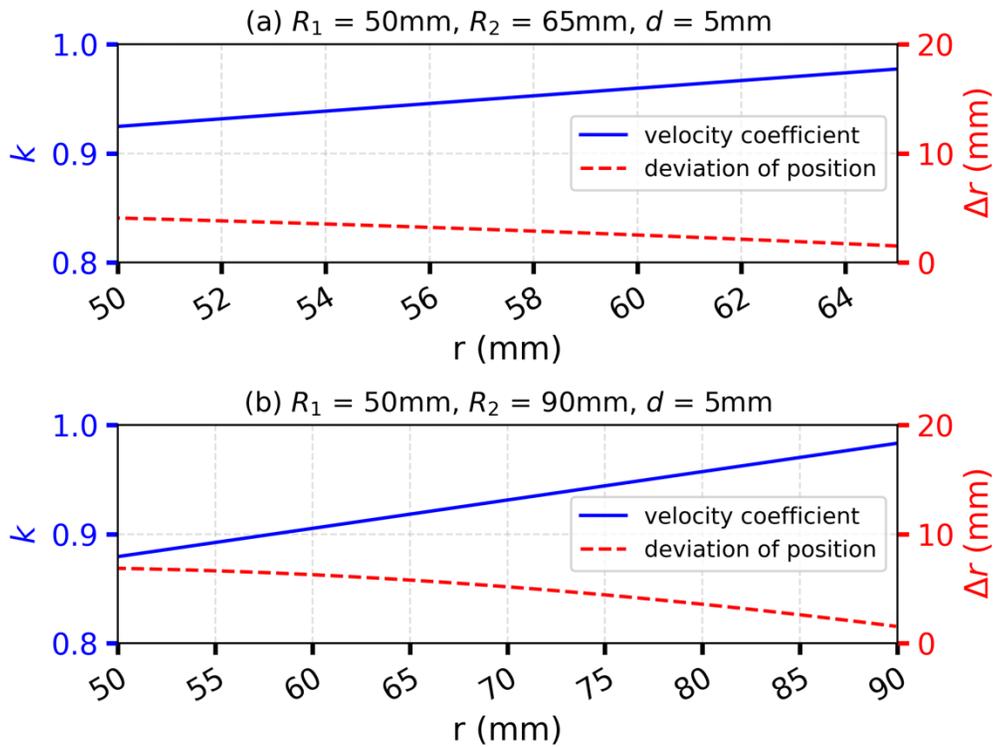

**MM Figure 2.** The correction parameters $k$ and $\Delta r = r_A - r_B$ changing with the radial position $r_B$.

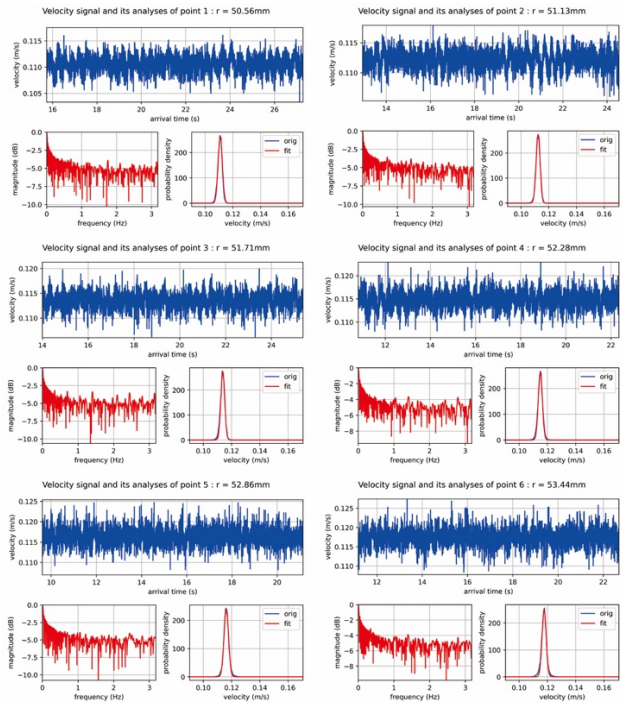

**MM Figure 3.1.** The velocity signal segment, the magnitude spectrum, the probability distribution and its Gaussian peak extraction for points 1-6, at the middle cross section with $R_2 = 65\text{mm}, \Omega = 20.88\text{ rpm}.$

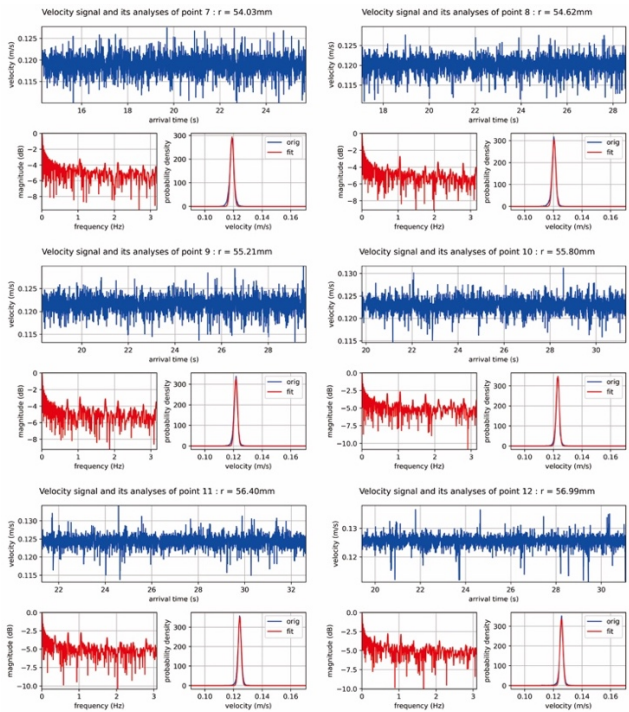

**MM Figure 3.2.** The velocity signal segment, the magnitude spectrum, the probability distribution and its Gaussian peak extraction for points 7-12, at the middle cross section with $R_2 = 65\text{mm}, \Omega = 20.88\text{ rpm}.$

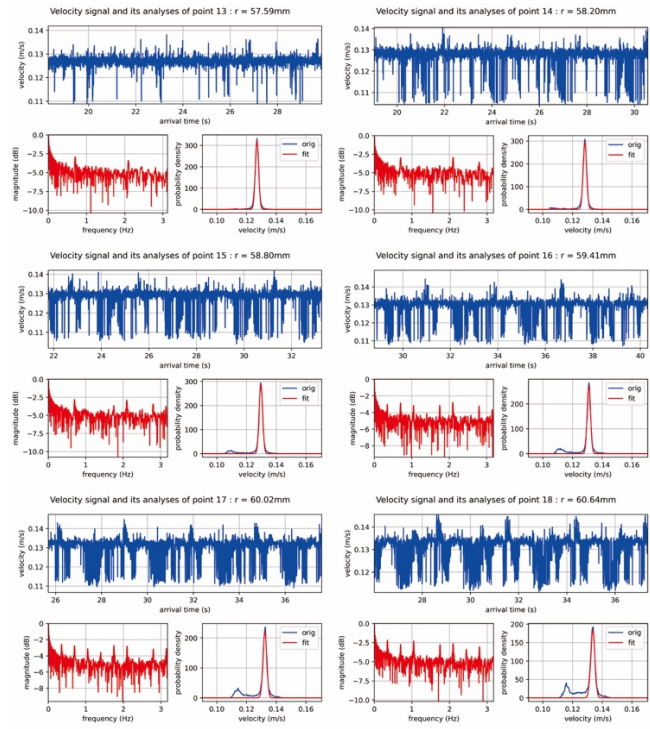

**MM Figure 3.3.** The velocity signal segment, the magnitude spectrum, the probability distribution and its Gaussian peak extraction for points 13-18, at the middle cross section with $R_2 = 65\text{mm}, \Omega = 20.88\text{ rpm}$.

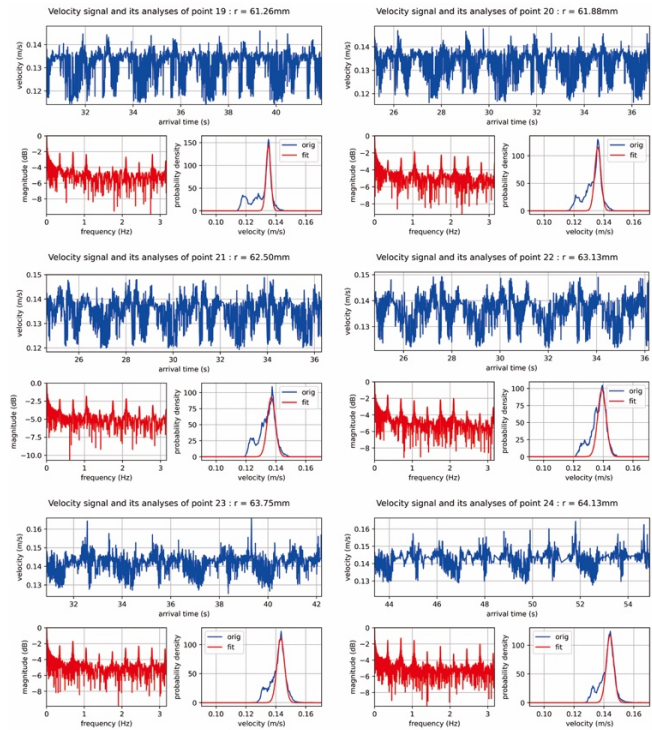

**MM Figure 3.4.** The velocity signal segment, the magnitude spectrum, the probability distribution and its Gaussian peak extraction for points 19-24, at the middle cross section with $R_2 = 65\text{mm}, \Omega = 20.88\text{ rpm}$.

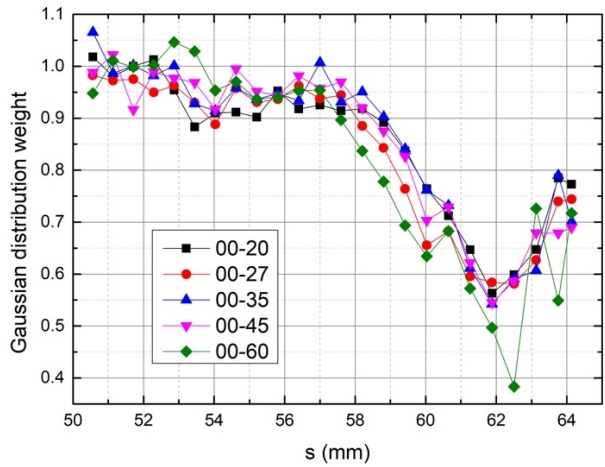

**MM Figure 4.** Weights of the extracted narrower Gaussian distribution of 5 groups of speed signals at different measurement points with $R_2 = 65\text{mm}$ and the measurement begin at 1 hour after startup. In the legend, "*a-b*" means that the rotation speed is about *b*rpm and the cross section is at *a*mm above the middle cross section.

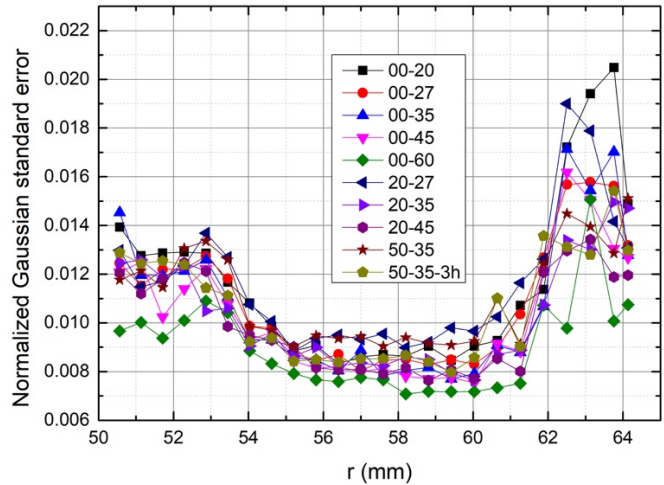

**MM Figure 5.** Normalized standard deviation of narrower Gaussian peak of 10 groups of speed signals with different rotating speeds, different axial sections and different stability times when $R_2 = 65\text{mm}$. In the legend, "*a-b*" means that the rotation speed is about *b*rpm and the cross section is at *a*mm above the middle cross section; "3h" means that the measurement begin at 3 hours after startup.

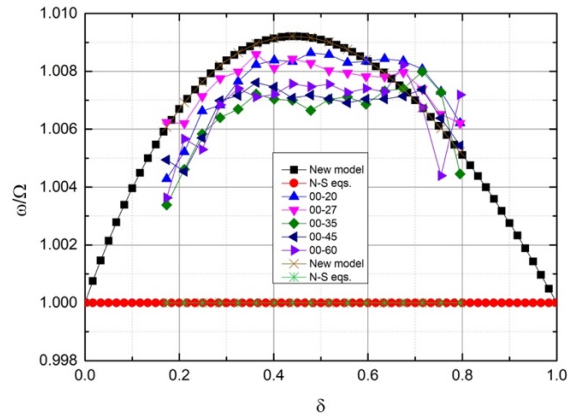

**MM Figure 6.** Normalized orbital angular speed of 5 groups of velocity information at the measurement points 5-21 extracted from the principal Gaussian distribution, and its comparison with the solutions of the N-S equations and new model when $R_2 = 65\text{mm}$. In the legend, "*a-b*" means that the rotation speed is about *b*rpm and the cross section is at *a*mm above the middle cross section.

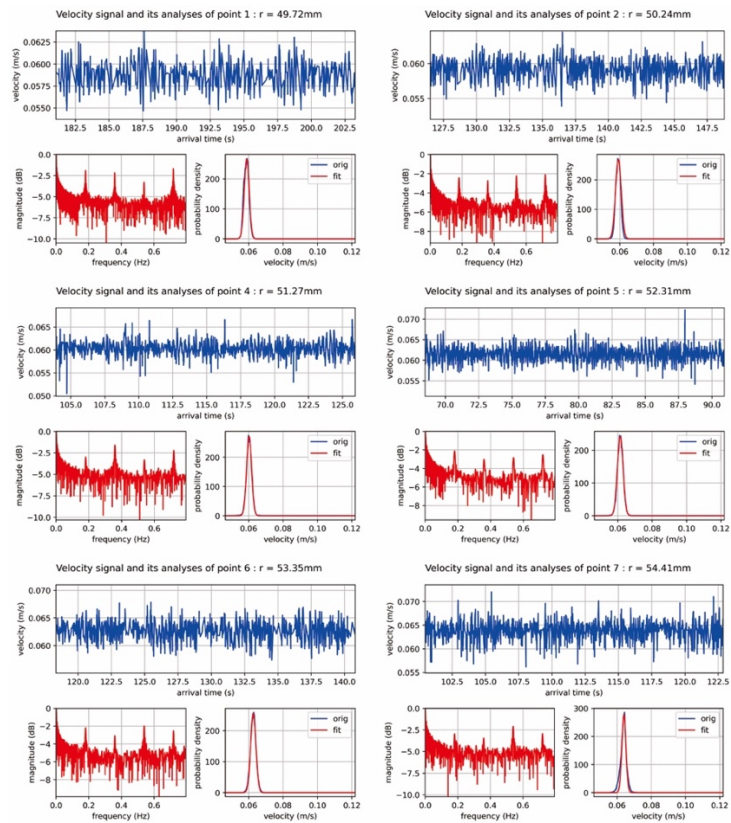

**MM Figure 7.1.** The velocity signal segment, the magnitude spectrum, the probability distribution and its Gaussian peak extraction for points 1, 2, 4-7 when $R_2 = 90\text{mm}, \Omega = 10.74 \text{ rpm}$.

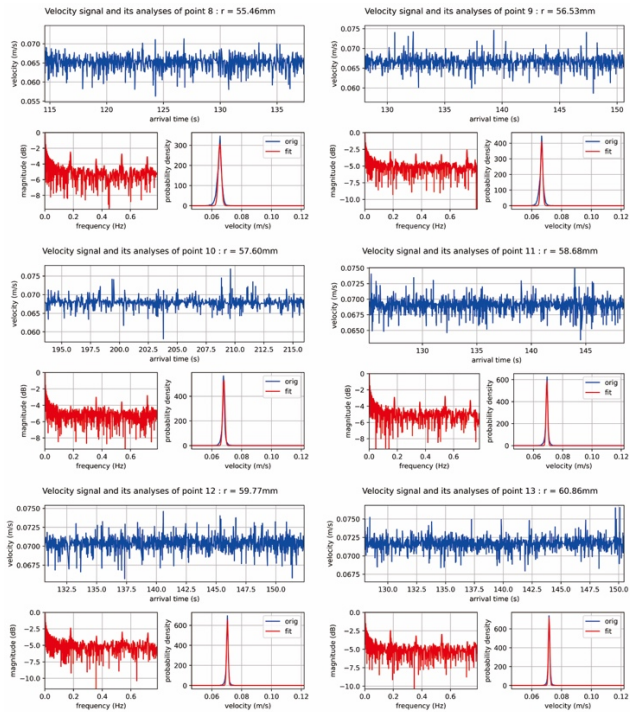

**MM Figure 7.2.** The velocity signal segment, the magnitude spectrum, the probability distribution and its Gaussian peak extraction for points 8-13, at the middle cross section with $R_2 = 90\text{mm}, \Omega = 10.74 \text{ rpm}$.

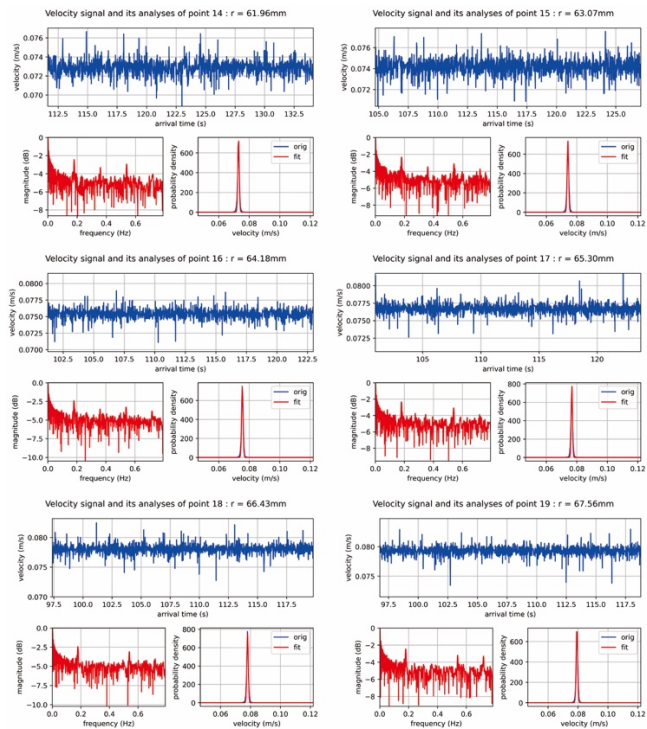

**MM Figure 7.3.** The velocity signal segment, the magnitude spectrum, the probability distribution and its Gaussian peak extraction for points 14-19, at the middle cross section with $R_2 = 90\text{mm}, \Omega = 10.74 \text{ rpm}$.

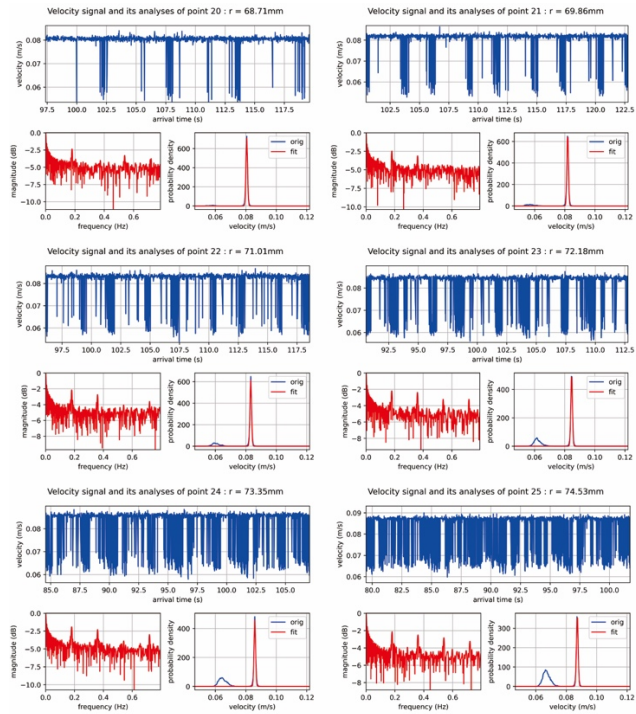

**MM Figure 7.4.** The velocity signal segment, the magnitude spectrum, the probability distribution and its Gaussian extraction for points 20-25, at the middle cross section with $R_2 = 90\text{mm}, \Omega = 10.74 \text{ rpm}.$

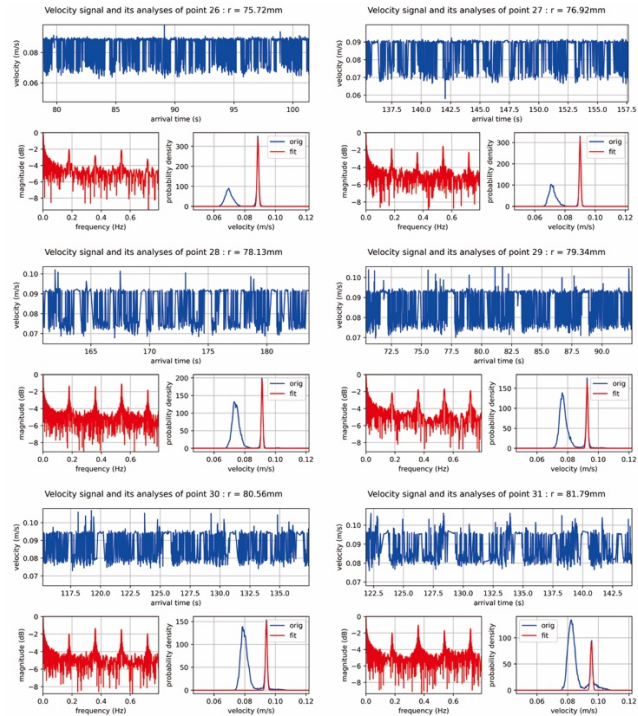

**MM Figure 7.5.** The velocity signal segment, the magnitude spectrum, the probability distribution and its Gaussian peak extraction for points 26-31, at the middle cross section with $R_2 = 90\text{mm}, \Omega = 10.74 \text{ rpm}.$

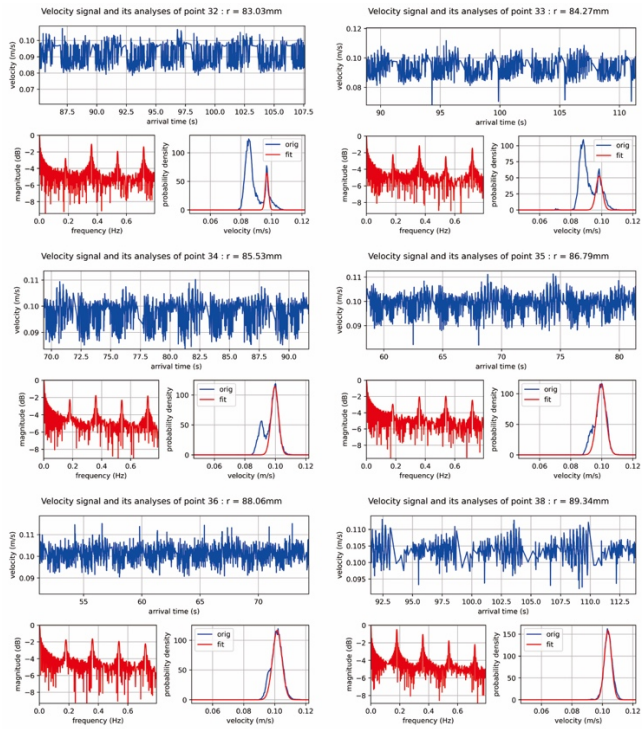

**MM Figure 7.6.** The velocity signal segment, the magnitude spectrum, the probability distribution and its Gaussian peak extraction for points 32-36, 38, at the middle cross section with $R_2 = 90\text{mm}, \Omega = 10.74\text{ rpm}$.

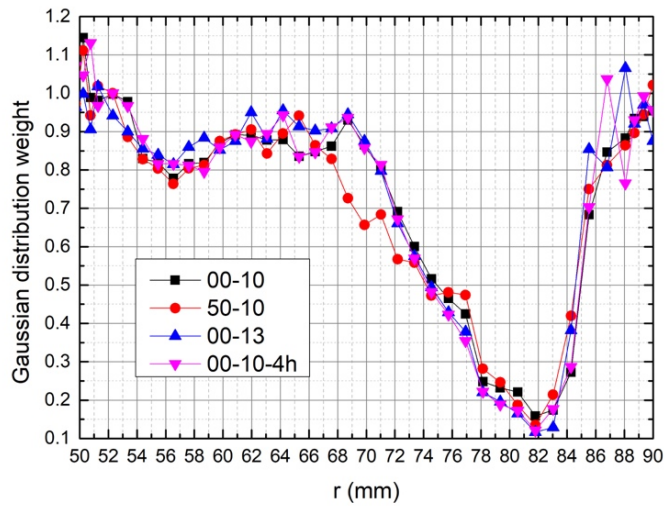

**MM Figure 8.** Weights of the extracted narrower Gaussian distribution of 4 groups of speed signals at different measurement points at different rotating speeds when $R_2 = 90\text{mm}$. In the legend, "*a-b*" means that the rotation speed is about *b*rpm and the cross section is at *a*mm above the middle cross section; "4h" means that the measurement begin at 4 hours after startup.

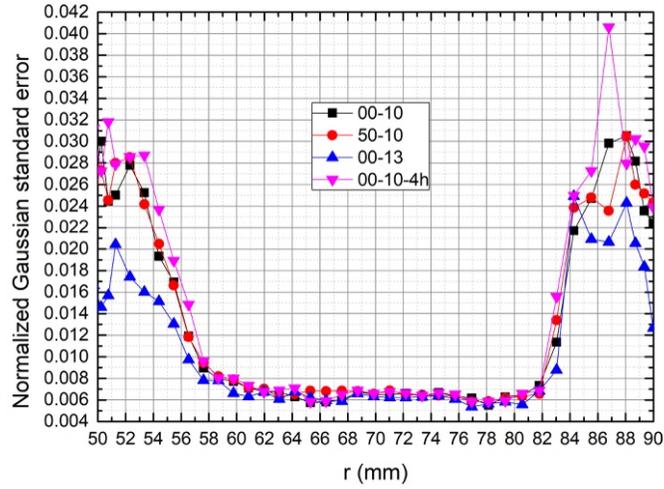

**MM Figure 9.** Normalized standard deviation of narrower Gaussian peak of 4 groups of speed signals with different rotating speeds, different axial sections and different stability times when $R_2 = 90\text{mm}$. In the legend, "*a-b*" means that the rotation speed is about *b*rpm and the cross section is at *a*mm above the middle cross section; "4h" means that the measurement begin at 4 hours after startup.

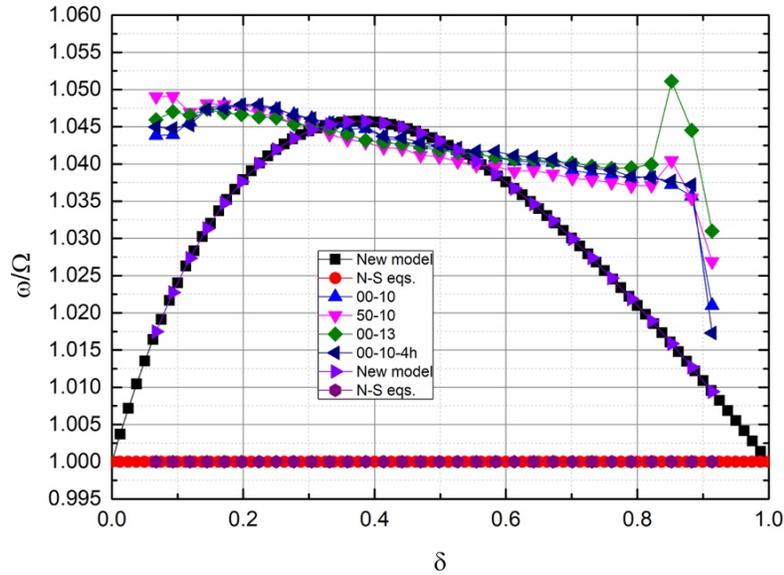

**MM Figure 10.** Normalized orbital angular speed of 4 groups of velocity information at the measurement points 5-35 extracted from the narrower Gaussian distribution, and its comparison with the solutions of the N-S equations and new model when $R_2 = 90\text{mm}$. In the legend, "*a-b*" means that the rotation speed is about *b*rpm and the cross section is at *a*mm above the middle cross section; "4h" means that the measurement begin at 4 hours after startup.

| LDA | First-dimension lasers | Second-dimension lasers |
|---|---|---|
| Wavelength [nm] | 660 (Red) | 785 (Infrared) |
| Beam diameter [mm] | 2.6 | |
| Beam spacing [mm] | 60 | |
| Probe focal distance [mm] | 300 | |
| Number of fringes | 29 | |
| Fringe spacing [μm] | 3.287 | 3.963 |
| Beam half-angle [°] | 5.711 | |
| Probe volume dX [mm] | 0.09745 | 0.1159 |
| Probe volume dY [mm] | 0.09696 | 0.1153 |
| Probe volume dZ [mm] | 0.9745 | 1.159 |

**MM Table 1.** Parameters of LDA measurement system.

| | | Middle CS | | 20mm high CS | | 50mm high CS | | 100mm high CS | |
|---|---|---|---|---|---|---|---|---|---|
| | | mean | min/max | mean | min/max | mean | min/max | mean | min/max |
| **IC** | $\Delta D_1$ | $-0.858\overline{3}$ | −0.92/−0.78 | $-0.85\overline{3}$ | −0.88/−0.78 | −0.86 | −0.92/−0.80 | $-0.818\overline{3}$ | −0.86/−0.78 |
| **OC1** | $\Delta(D_2 + 2d)$ | $0.1691\dot{6}$ | −0.18/0.50 | $0.31\dot{6}$ | −0.04/0.80 | 0.305 | −0.18/0.78 | - | - |
| | $\Delta d$ | $0.0841\dot{6}$ | −0.19/0.50 | 0.1325 | −0.11/0.52 | 0.1396 | −0.08/0.48 | - | - |
| **OC2** | $\Delta(D_2 + 2d)$ | $0.1808\dot{3}$ | −0.28/0.60 | - | - | $0.0008\dot{3}$ | −0.44/0.92 | $0.0191\dot{6}$ | −0.60/0.86 |
| | $\Delta d$ | −0.2171 | −0.44/0.07 | - | - | $-0.24\dot{3}$ | −0.52/0.10 | −0.1854 | −0.50/0.23 |

Note: 'CS' - cross section, 'IC' - inner cylinder, 'OC' - outer cylinder.

**MM Table 2.** Deviations (unit: mm) from the expected geometry of cylinders due to machining difficulty, where the expected geometrical parameters $\{R_1, R_2, d\}$ of IC/OC1 and IC/OC2 are $\{50, 65, 5\}$ and $\{50, 90, 5\}$, respectively.